\documentclass[aps,prl,groupedaddress,showpacs,twocolumn]{revtex4-1}
\usepackage{graphicx}

\begin{document}

\title{Collective Motion of Moshers at Heavy Metal Concerts}

\author{Jesse L. Silverberg}
\email[]{JLS533@cornell.edu}
\author{Matthew Bierbaum}
\author{James P. Sethna}
\author{Itai Cohen}
\affiliation{Laboratory of Atomic and Solid State Physics and Department of Physics, Cornell University, Ithaca NY 14853, USA}

\date{\today}

\begin{abstract}
Human collective behavior can vary from calm to panicked depending on social context.  Using videos publicly available online, we study the highly energized collective motion of attendees at heavy metal concerts.  We find these extreme social gatherings generate similarly extreme behaviors: a disordered gas-like state called a \textit{mosh pit} and an ordered vortex-like state called a \textit{circle pit}.  Both phenomena are reproduced in flocking simulations demonstrating that human collective behavior is consistent with the predictions of simplified models.
\end{abstract}


\maketitle

\begin{figure}
\centering
\scalebox {.8}{\includegraphics{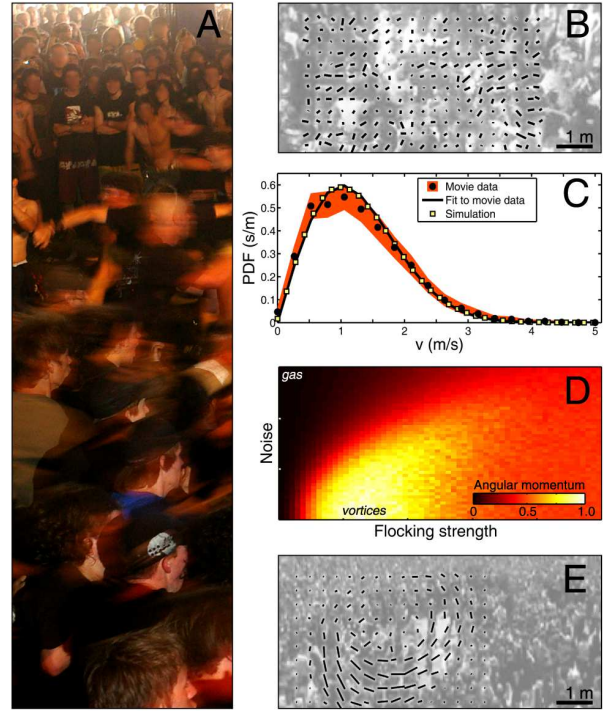}}
\caption{(A) Example image of crowd at a heavy metal concert.  (B) Single video frame illustrating a characteristic mosh pit with overlaid velocity field.  (C) Speed probability distribution function (PDF) for the movie in (B) (circles), the best fit to the 2D Maxwell-Boltzmann speed distribution (solid), and simulated speed distribution (squares).  (D) Simulated phase diagram plotting the MASHer RMS angular momentum, demonstrating the existence of mosh pits (gas) and circle pits (vortices).  (E) Single video frame illustrating a characteristic circle pit with overlaid velocity field.}
\label{fig1}
\end{figure}

Human collective behaviors vary considerably with social context.  For example, lane formation in pedestrian traffic \cite{Moussaid:2011}, jamming during escape panic \cite{Helbing:2007}, and Mexican waves at sporting events \cite{Farkas:2002} are emergent phenomena that have been observed in specific social settings.  Here, we study large crowds ($10^2 - 10^5$ attendees) of people under the extreme conditions typically found at heavy metal concerts.  Often resulting in injuries \cite{Janchar:2000}, the collective mood is influenced by the combination of loud, fast music (130 dB \cite{Drake-Lee:1992}, 350 beats per minute), synchronized with bright, flashing lights, and frequent intoxication \cite{Lim:2008}.  This variety and magnitude of stimuli are atypical of more moderate settings, and contribute to the collective behaviors studied here (Fig.~\ref{fig1}(A)).

Videos filmed by attendees at heavy metal concerts \cite{YT1} highlight a collective phenomenon consisting of $10^1 - 10^2$ participants commonly referred to as a \textit{mosh pit}.  In mosh pits, the participants (moshers) move randomly, colliding with one another in an undirected fashion (Fig.~\ref{fig1}(B)).  Qualitatively, this phenomenon resembles the kinetics of gaseous particles, even though moshers are self-propelled agents that experience dissipative collisions.  To explore this analogy quantitatively, we obtained video footage, corrected for perspective distortions \cite{Py:2005} as well as camera instability, and used PIV analysis \cite{Sveen:2003} to measure the two-dimensional (2D) velocity field on an interpolated grid.  From this data, we calculated the velocity-velocity correlation function $c_{vv}$ and noted an absence of the spatial oscillations typically found in liquid-like systems.  Additionally, $c_{vv}$ was well fit by a pure exponential ($R^2 = 0.97$) with a decay length of 0.78 m.  Taken together, these findings offer strong support for the analogy between mosh pits and gases.  As a further check, we examined the 2D speed distribution; previous observations of human pedestrian traffic and escape panic led us to expect a broad distribution not well described by simple analytic expressions \cite{Henderson:1971,Helbing:2007}.  However, the measured speed distribution in mosh pits was well fit by the equilibrium speed distribution of classical 2D gasses (Fig.~\ref{fig1}(C)), otherwise known as the Maxwell-Boltzmann distribution \cite{Landau:1980}.  These observations present an interesting question: Why does an inherently non-equilibrium system exhibit equilibrium characteristics?  

Studies of collective motion in living and complex systems have found notable success within the framework of flocking simulations \cite{Couzin:2002,Czirok:1996,Schaller:2010,Leonard:2012,Vicsek:2012}.  Thus, we use a Vicsek-like model \cite{Vicsek:1995} to simplify the complex behavioral dynamics of each human mosher to that of a simple soft-bodied particle we dub a Mobile Active Simulated Humanoid, or MASHer (SI).  Our model includes two species of MASHers to reflect the typical crowd at heavy metal concerts (Fig.~\ref{fig1}(A)) \cite{You:2009}.  Active MASHers are self-propelled, experience flocking interactions, and are subject to random fluctuations in the forces they experience.  Passive MASHers prefer to remain stationary, and are not subject to flocking interactions or random forces.  Though initially both populations are uniformly mixed, we found that with sufficient time the flocking-interaction leads to a spontaneous phase separation with a dense population of active MASHers confined by passive MASHers.  When noise dominates the flocking and propulsion terms, the motion of active MASHers resembles the motion in actual mosh pits and quantitatively reproduces the experimentally observed statistics (Fig.~\ref{fig1}(C)).  These results demonstrate how a non-equilibrium system can have equilibrium characteristics: random motions over a sufficient time reproduce the statistics of classical gasses via the Central Limit Theorem.  

Conversely, when the flocking term dominates active MASHer motion, our model predicts a highly ordered vortex-like state \cite{Bazazi:2012} where MASHers again phase separate, but the confined active MASHers move with a large non-zero angular momentum (Fig.~\ref{fig1}(D)).  Remarkably, this spontaneous phase separation and vortex formation is also observed at heavy metal concerts where they are conventionally called \textit{circle pits} (Fig.~\ref{fig1}(E)) \cite{YT1}.  In simulations, we found an even distribution between clockwise (CW) and counter-clockwise (CCW) motion when viewed from above, whereas our observations from concerts show 5\% flow CW with the remaining 95\% flowing CCW ($p < 0.001$).  This asymmetry is independent of geographical location, as video data was collected from a variety of countries including the United State of America, the United Kingdom, and Australia.  Though the origin of this effect is unknown, we speculate it may be related to the dominant handedness/footedness found in humans.

The collective behavior described here has not been predicted on the basis of staged experiments with humans \cite{Isobe:2004,Moussaid:2009}, making heavy metal concerts a unique model system for reliably, consistently, and ethically studying human collective motion.  Further studies in this unique environment may enhance our understanding of collective motion in riots, protests, and panicked crowds, leading to new architectural safety design principles that limit the risk of injury at extreme social gatherings.  Moreover, these concerts have the further advantage of exhibiting a rich variety of states such as the \textit{wall of death} (moshers split into two groups separated by an open space and, when signaled, simultaneously run at the opposing group), collective jumping (locally correlated, globally decorrelated), and soft modes \cite{Silbert:2009} in jammed attendees.  Thus, the extreme conditions at heavy metal concerts offer new opportunities for studying a wide range of collective behaviors that arise in extreme social gatherings of large human groups.

\appendix*
\section{Supplemental Information}
\paragraph{Modeling Moshers with MASHers} To model the crowd at heavy metal concerts, we simplify the complex behavioral dynamics of each human \textit{mosher} to that of a simple particle we dub a Mobile Active Simulated Humanoid, or MASHer.  Analogous to previous work \cite{Vicsek:1995,Helbing:1995}, we include the following ingredients in our model: (1) Hertzian soft body repulsion (2) self-propulsion (3) flocking interactions and (4) Gaussian random noise.  The forces on MASHers $i$ are:
\begin{eqnarray}
\vec{F}_{i}^{\rm repulsion} & = & \left\{ \begin{array}{ll}
\epsilon \left(1-\frac{r_{ij}}{2r_0}\right)^{5/2} \hat{r}_{ij} \quad & {\rm for} \quad r_{ij} < 2r_0 \\
0 & {\rm otherwise} \\
\end{array} \right., \nonumber \\
\vec{F}_{i}^{\rm propulsion} & = & \mu (v_0 - v_i) \hat{v}_i, \nonumber \\
\vec{F}_{i}^{\rm flocking} & = & \alpha \sum_{j=0}^{N_i} \vec{v}_j \Big/ \left|\sum_{j=0}^{N_i} \vec{v}_j \right|, \nonumber \\
\vec{F}_{i}^{\rm noise} & = & \vec{\eta}_i. 
\label{modeleq}
\end{eqnarray}
Here, $r_{ij}$ is the center-to-center distance between MASHer $i$ and $j$, $r_0$ is the MASHer radius, $v_0$ is its preferred speed, and $v_i$ is its instantaneous speed.   The coefficients $\epsilon$, $\mu$, and $\alpha$ are free parameters in our model reflecting the relative strength of each term, and the carat (~$\hat{}$~) denotes unit vectors.  The flocking force is a summation of the velocities $v_j$ of all the $N_i$ MASHers within a radius $r_{\rm flock}$ of MASHer $i$.  $\vec{\eta}_i$ is a vector whose component are drawn from a Gaussian distribution with zero mean and whose standard deviation was varied in the simulation.  In total we have $7$ independent parameters, of which all but two are held constant.

At heavy metal concerts, there are people who do and do not participate in moshing.  We reflect this in our model by having two species.  For active MASHers, we have, in the appropriate units, $v_0 = 1, \alpha$ varies over the interval $[0,5],$ and the standard deviation of the components of $\vec{\eta}_i$ varies over the interval $[0,3]$.  For passive MASHers $v_0 = \alpha = |\vec{\eta}_i| = 0$.  By varying the magnitude of the flocking coefficient and the strength of the noise for active MASHers, we investigate the phase diagram of the model.  The remaining parameters are identical between the two populations, $\epsilon = 100, \mu = 0.05, r_0 = 1$, and $r_{\rm flock} = 4$.  

\paragraph{Simulations} We simulate concerts with $N = 500$ attendees, of which $N_a = 150$ are active MASHers and remaining are passive MASHers.  Initially, the active MASHers are situated in a circle centered in the middle of the simulation box.  This choice of initial positions is a choice of convenience: any non-zero flocking force causes MASHers to self-segregate when started with random initial positions, though this tends to have a large timescale ($\sim 10^3 \times (r_0/v_0)$ time steps) when $r_{\rm flock}/r_0$ is ${\cal O}(1)$.  Once formed, simulated mosh and circle pits typically last for times greater than $10^5 \times (r_0/v_0)$ time steps, and can switch directions at random intervals \cite{Chen:2012}.  The simulation box is of length $L = 1.03\sqrt{\pi r_0^2 N} \approx 40.8$ to give a packing fraction $\rho = 0.94$, and is periodic on both sides to avoid boundary effects.  We use the Newton-Stomer-Verlet integration algorithm and cell based neighbor lists to expedite computation.  The phase diagram shown in the main text is comprised of $2 \times 10^6$ individual simulations sampled on $10,000$ grid points.  For each run, we measured the angular momentum about the center of mass $x_{cm} = (L / 2\pi) \rm{arctan}( Im(A) / Re(A) )$, where $A = \sum_{i=0}^{N_a-1} \exp\left(- 2\pi i x_i / L \right)$, with a similar expression for $y_{cm}$. 

Our flocking simulation has three time scales defined by the model parameters: the self-propulsion time scale $\tau_{\rm prop} = 1 / \mu$, the flocking time scale $\tau_{\rm flock} = v_0 / \alpha$, and the collision time scale $\tau_{\rm coll} = 1/(2r_0 v_0 \rho)$.  Both collisions and noise tend to randomize motion, whereas flocking and self-propulsion tend to homogenize motion.  Thus, when the time between random collisions $\tau_{\rm coll}$ is much less than $\tau_{\rm prop}$ and $\tau_{\rm flock}$, the statistical motion of the system is described by a single effective temperature and we recover the Maxwell-Boltzmann speed distribution reported in the main text.

\begin{figure}
\centering
\scalebox{1}{\includegraphics{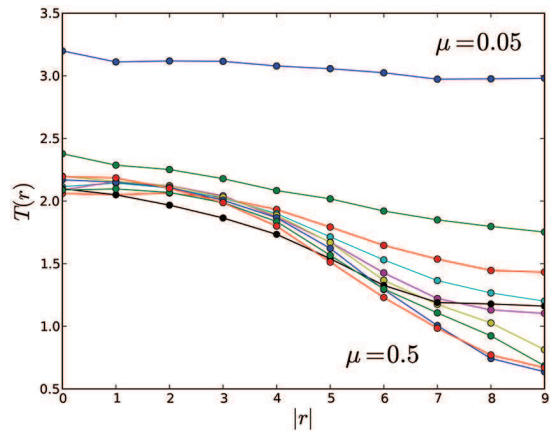}}
\caption{ As the self-propulsion coefficient $\mu$ increases, we find the temperature $T(r)$ has a radial dependence in simulated moshpits.  This arises from collisions between active and passive MASHers that remove energy from the boundary faster than collisions among active MASHers can rethermalize the system.  Each line corresponds to values of $\mu$ starting from 0.05 and increasing to 0.50 in steps of 0.05.  Simulations described in the main text correspond to $\mu = 0.05$ and have essentially a single effective temperature independent of radius.}
\label{fig2}
\end{figure}

If we increase the self-propulsion coefficient $\mu$ (decrease $\tau_{\rm prop}$), we find the motion, though random, is no longer fit by a Maxwell-Boltzmann distribution.  Instead, collisions between active and passive MASHers on the boundary of the simulated mosh pit removes energy faster than collisions among active MASHers can rethermalize the system.  Consequently, measurements \textit{in silico} show a radial temperature gradient is established with a higher effective temperature at the core of the simulated mosh pit and a lower effective temperature at the edge (Fig.~\ref{fig2}).  The speed distribution in the limit of small $\tau_{\rm prop}$ is therefore a sum of Maxwell-Boltzmann distributions characterized by different temperatures.

\paragraph*{Source Code} Source code and a phase diagram generating Python script are available under the M.I.T. license on github.com at \url{https://github.com/mattbierbaum/moshpits}.  An interactive Javascript version of the simulation is available at
\url{http://mattbierbaum.github.com/moshpits.js}.

\begin{acknowledgments}
The photo in Fig.~\ref{fig1}(A) was taken and graciously provided by Ulrike Biets.  JLS and MB also thank D. Porter, L. Ristroph, J. Freund, J. Mergo, A. Holmes, A. Alemi, M. Flashman, K. Prabhakara, J. Wang, R. Lovelace, P. McEuen and S. Strogatz.  Fieldwork was independently funded by JLS.
\end{acknowledgments}

\end{document}